\documentclass[authoryear]{elsarticle}

\usepackage{hyperref}
\usepackage{lineno}
\modulolinenumbers[5]
\usepackage{bm}
\usepackage{amsmath}
\usepackage{amssymb}
\usepackage{subfigure}
\usepackage{color}
\usepackage{graphicx}
\usepackage{amssymb}

\journal{Planetary and Space Science}

\begin{document}

\begin{frontmatter}

  \title{High-Resolution Simulations of Catastrophic Disruptions: Resultant Shape Distributions}

\author[mymainaddress]{Keisuke Sugiura\corref{mycorrespondingauthor}}
\cortext[mycorrespondingauthor]{Corresponding author}
\ead{sugiuraks@elsi.jp}

\author[mysecondaddress]{Hiroshi Kobayashi}

\author[mysecondaddress]{Shu-ichiro Inutsuka}

\address[mymainaddress]{Earth-Life Science Institute, Tokyo Institute of Technology, Tokyo 152-8550, Japan}
\address[mysecondaddress]{Department of Physics, Nagoya University, Aichi 464-8602, Japan}

\begin{abstract}
  The members of asteroid families have various shapes. We investigate the origin of their shapes by high-resolution impact simulations for catastrophic disruptions using a Smoothed Particle Hydrodynamics code. Collisional remnants produced through our simulations of the catastrophic disruptions mainly have spherical or bilobed shapes. However, no flat remnants with the ratio of minor to major axis lengths $c/a \lesssim 0.5$ are formed. The results of the simulations provide various shapes of asteroids and explain most of the shapes in asteroid families that are supposed to be produced through catastrophic disruptions. However, the present simulations do not explain significantly flat asteroids. We suggest that these flat asteroids may be interlopers or formed through low-velocity collisions between member asteroids.
\end{abstract}


\end{frontmatter}

\section{Introduction}
Asteroids in the present solar system are considered to be remnant planetesimals that existed in the planet formation era (e.g.,\,\citealt{Petit-et-al2001, Bottke-et-al2005, Kobayashi-et-al2016}). Especially, asteroids with diameters larger than $\sim 10\,{\rm km}$ may keep their shapes formed in the primordial environment because the occurrence timescales of destructive collisions (collisional lifetimes) of those asteroids are estimated to be comparable to or much longer than the age of the solar system (\citealt{Obrien-and-Greenberg2005}). On the other hand, the existence of asteroid families implies recent impacts between asteroids (e.g.,\,\citealt{Hirayama1918}), which alter the shapes of asteroids. Thus the shapes of large asteroids may tell us information about collisions that occurred in the distant or recent past.

Owing to material strengths and small sizes of asteroids, they keep a variety of shapes including irregular shapes different from spherical shapes (e.g.,\citealt{Cibulkova-et-al2016}). Shapes of asteroids are estimated or clarified through light-curve (e.g.,\,\citealt{Kaasalainen-et-al2003}), radar (e.g.,\,\citealt{Shepard-et-al2017}), and in-situ (e.g.,\,\citealt{Fujiwara-et-al2006}) observations. The Database of Asteroid Models from Inversion Techniques (DAMIT: \citealt{Durech-et-al2010}) provides shape models of about 1,600 asteroids obtained from light-curve observations. Thus, the shape distribution is one of important statistical information for asteroids.

Shapes of asteroids are mainly formed through asteroidal collisions, and resultant shapes produced through impacts depend on impact conditions such as impact velocities (e.g.,\,\citealt{Leinhardt-et-al2000, Jutzi-and-Asphaug2015, Sugiura-et-al2018}). The average impact velocity in the present main belt is estimated to be about $5\,{\rm km/s}$ (e.g.,\,\citealt{OBrien-and-Sykes2011, Sugiura-et-al2018}), so that impacts between asteroids mainly result in catastrophic disruptions. On the other hand, collisional growth of planetesimals may form the size distribution of $100\,{\rm km}$ sized main-belt asteroids (\citealt{Kobayashi-et-al2016}). The impact velocities in such formation era do not significantly exceed their escape velocities, so that impacts mainly result in non-destructive or merging collisions. Therefore, understanding of shapes produced through non-destructive impacts and catastrophic disruptions may provide valuable information on the formation era of each asteroid.

\cite{Sugiura-et-al2018} conducted numerical simulations to investigate asteroidal shapes produced through non-destructive and equal-mass impacts and showed that the impacts produce various shapes including extremely flat and elongated shapes. However, they investigate only the largest remnants produced through non-destructive impacts because the numerical resolution is relatively limited, and shapes of remnants produced through catastrophic disruption are not investigated. \cite{Schwartz-et-al2018} found that bilobed shapes such as the shape of the comet 67P/Churyumov-Gerasimenko are formed through catastrophic disruptions. 

Asteroid families are groups of asteroids that have similar proper orbital elements, and a family is considered to be formed by the destruction of single parent body (\citealt{Hirayama1918, Farinella-et-al1996}). The ratio of the mass of the largest remnant in a family to the mass of all member asteroids (or the mass of a parent body) tells us how destructive a family forming impact was. Some families such as the Eos family were formed through catastrophic disruptions (\citealt{Broz-et-al2013}). Thus the comparison between shapes of family asteroids and shapes of collisional outcomes obtained from numerical simulations may provide some insights into the physics of catastrophic disruptions or the evolution of asteroid families.

In this study, we investigate the shapes of not only the largest but also smaller remnants produced by catastrophic disruptions through high-resolution simulations. We use a Smoothed Particle Hydrodynamics (SPH) code throughout simulations although previous studies use $N$-body codes for reaccumulation phases (e.g.,\,\citealt{Michel-and-Richardson2013, Schwartz-et-al2018}). We compare shapes of remnants obtained from our simulations with those of asteroids belonging to families produced through catastrophic disruptions.

\section{Method, initial conditions, and analysis of results}
For numerical simulations of catastrophic collisions, we use the SPH method for elastic dynamics (\citealt{Libersky-and-Petschek1991}). We utilize the same version of a SPH code described in \cite{Sugiura-et-al2018}. We use a pressure dependent yielding criterion (\citealt{Jutzi2015}) and a fracture model (Grady-Kipp fragmentation model: \citealt{Benz-and-Asphaug1995}) for undamaged material, and use a friction model (Drucker-Prager yield criterion without cohesion: \citealt{Jutzi2015}) for completely damaged material. We also solve the self gravity between SPH particles. We use the Tillotson equation of state (\citealt{Tillotson1962}). The simulation code is parallelized using Framework for Developing Particle Simulator (\citealt{Iwasawa-et-al2015, Iwasawa-et-al2016}).

We use the parameter set of basaltic material (\citealt{Benz-and-Asphaug1999}) for the Tillotson equation of state and the fracture model. The uncompressed density in this parameter set is $2.7\,{\rm g/cm^{3}}$. We ignore the increase of the macroporosity of reaccumulated remnants, and thus the density of remnants is almost the same as the uncompressed density. The cohesion of undamaged material is set to $100\,{\rm MPa}$. The friction angle for completely damaged rocks is set to $40^{\circ}$. Regarding to other parameters, we use the same values as described in \cite{Sugiura-et-al2018}.

We investigate collisions between uniform, uncompressed, and intact basaltic spheres with zero rotation and radii $R_{{\rm t}}$ and $R_{{\rm i}}$. We set the radius of the target body $R_{{\rm t}}=50\,{\rm km}$. We conduct four simulations of catastrophic collisions with the different impact conditions: the ratio of the impactor mass to the target mass $q=(R_{{\rm i}}/R_{{\rm t}})^{3}$, the impact velocity $v_{{\rm imp}}$, and the impact angle $\theta_{{\rm imp}}$ ($\theta_{{\rm imp}}=0^{\circ}$ corresponds to a head-on collision). The detailed impact conditions are listed in Table \ref{tab1}. We use about $4\times 10^{6}$ SPH particles for each impact simulation.

\begin{table}[!htb]
  \begin{center}
    \begin{tabular}{c | c c c }\hline 
       & $q$ & $v_{{\rm imp}}$ (m/s) & $\theta_{{\rm imp}}$ ($^{\circ}$)\\
      \hline 
      Impact 1  & 1/1 & 350 & 15 \\ 
      Impact 2  & 1/4 & 700 & 15 \\
      Impact 3  & 1/8 & 1000 & 0 \\
      Impact 4  & 1/16 & 1700 & 15 \\
      \hline
    \end{tabular}
  \end{center}
  \caption{Impact parameters (colliders' mass ratio $q$, impact velocity $v_{{\rm imp}}$, and impact angle $\theta_{{\rm imp}}$) for impact simulations.}
  \label{tab1}
\end{table}

We continue the simulations of the impacts and subsequent reaccumulation of fragments until $1.0\times 10^{5}\,{\rm s}$ after the time of the impacts, which is much longer than the reaccumulation timescale estimated to be $2R_{{\rm t}}/v_{{\rm esc}} \approx 1,600\,{\rm s}$, where $v_{{\rm esc}}$ is the two-body escape velocity of the two impacting asteroids. Then, we find remnants produced through reaccumulation of fragments using a friend-of-friend algorithm (e.g.,\,\citealt{Huchra-and-Geller1982}). We identify swarms of SPH particles with spacing less than 1.5 smoothing lengths as remnants. Finally, we measure axis lengths of remnants composed of more than 5,000 SPH particles using a procedure described in \cite{Sugiura-et-al2018}. Note that a remnant with 5,000 SPH particles is resolved by about 20 SPH particles along each axis so that the shape of such a remnant is resolved to a certain extent. Two axis ratios $b/a$ and $c/a$ characterize shapes of remnants, where $a$, $b$, and $c$ represent the major, intermediate, and minor axis lengths, respectively.

\section{Results}

\begin{figure}[!htb]
 \begin{center}
 \includegraphics[bb=0 0 960 498, width=1.0\linewidth]{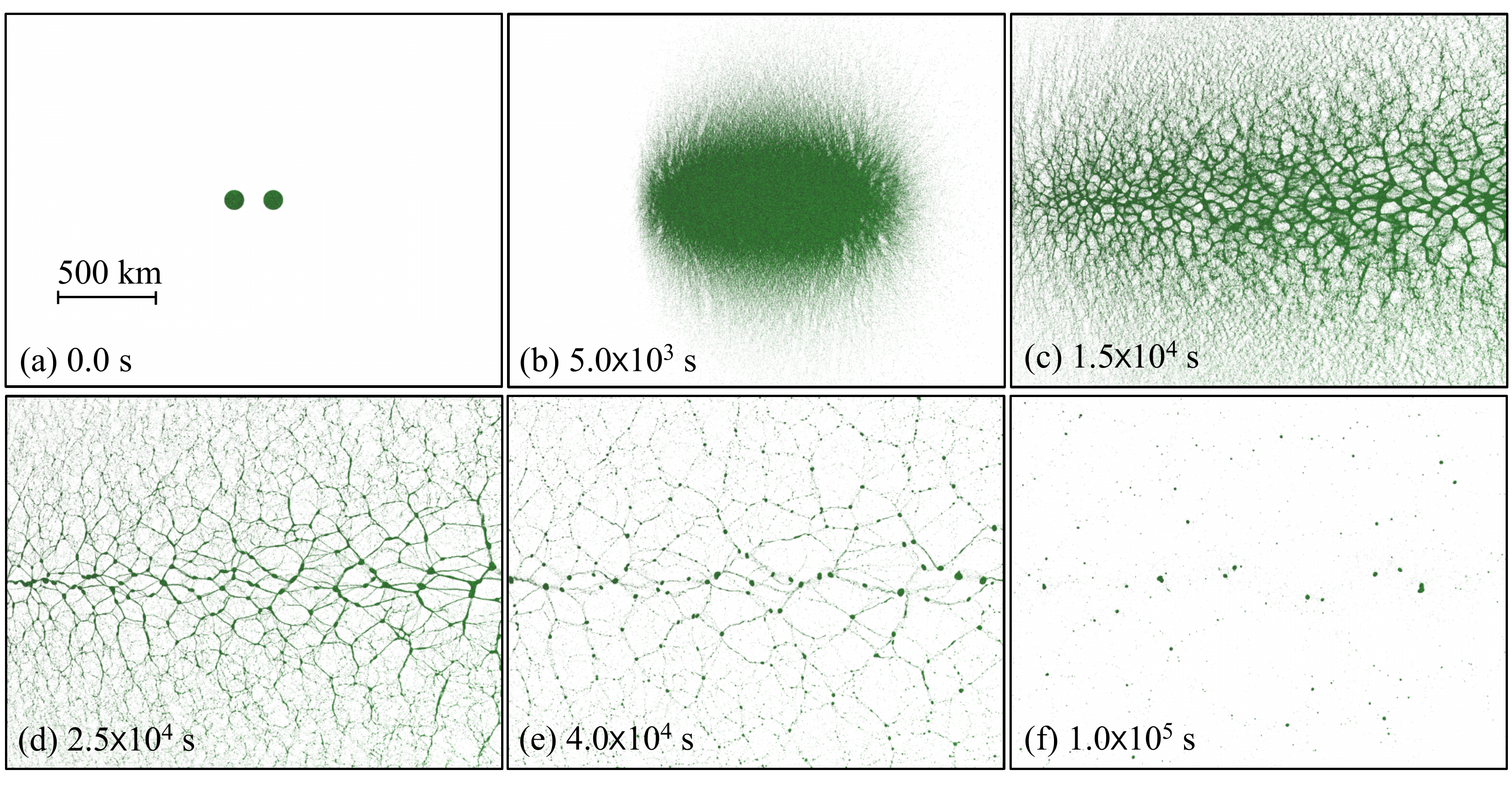}
 \caption{Snapshots of the simulation of Impact 1 ($q=1$, $v_{{\rm imp}}=350\,{\rm m/s}$, $\theta_{{\rm imp}}=15^{\circ}$). The snapshots show the face on views of the ejecta curtain produced through the impact.}
 \label{Rt=50km-q=1.0-theta=15-v=350ms-front-view} 
 \end{center}
\end{figure}

\begin{figure}[!htb]
 \begin{center}
 \includegraphics[bb=0 0 960 498, width=1.0\linewidth]{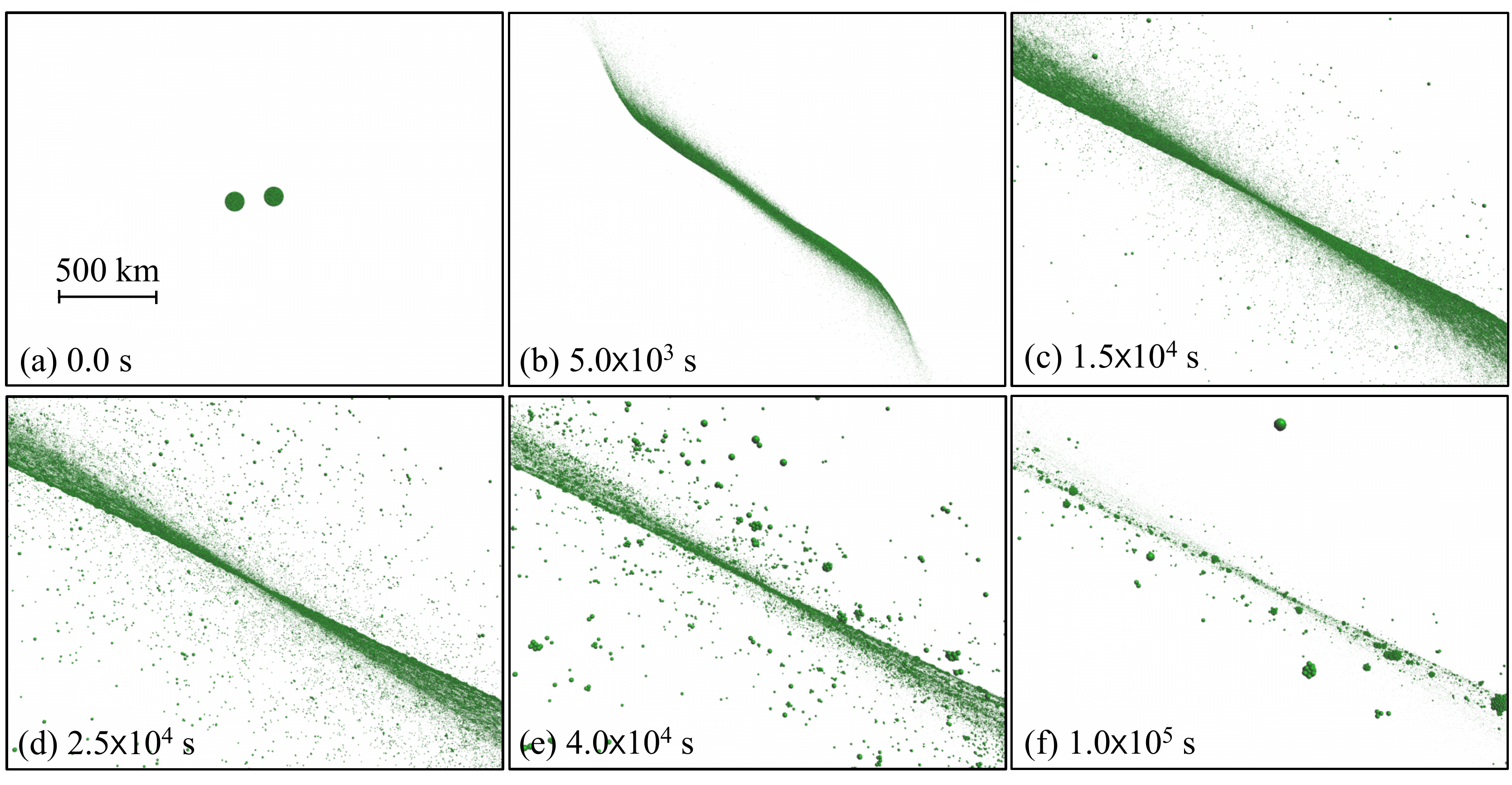}
 \caption{The same as Fig.\,\ref{Rt=50km-q=1.0-theta=15-v=350ms-front-view}, but these snapshots show the edge on views of the ejecta curtain produced through the impact.}
 \label{Rt=50km-q=1.0-theta=15-v=350ms-side-view} 
 \end{center}
\end{figure}

Figures \ref{Rt=50km-q=1.0-theta=15-v=350ms-front-view} and \ref{Rt=50km-q=1.0-theta=15-v=350ms-side-view} show the snapshots of the simulation of Impact 1 ($q=1$, $v_{{\rm imp}}=350\,{\rm m/s}$, $\theta_{{\rm imp}}=15^{\circ}$). The catastrophic disruption produces a wide and thin ejecta curtain (Figs.\,\ref{Rt=50km-q=1.0-theta=15-v=350ms-front-view}b and \ref{Rt=50km-q=1.0-theta=15-v=350ms-side-view}b). The gravitational instability and fragmentation of the curtain produce filamentary structures (Fig.\,\ref{Rt=50km-q=1.0-theta=15-v=350ms-front-view}d). The gravitational reaccumulation of fragments in the filamentary structures (Fig.\,\ref{Rt=50km-q=1.0-theta=15-v=350ms-front-view}e) eventually forms many small remnants (Figs.\,\ref{Rt=50km-q=1.0-theta=15-v=350ms-front-view}f and \ref{Rt=50km-q=1.0-theta=15-v=350ms-side-view}f). Note that these processes (the gravitational fragmentation of the sheets, the formation of the filaments, and the gravitational reaccumulation of fragments) are independent of the detailed impact conditions or initial particle placement as long as impacts result in catastrophic disruption.

\begin{figure}[!htb]
 \begin{center}
 \includegraphics[bb=0 0 370 252, width=1.0\linewidth]{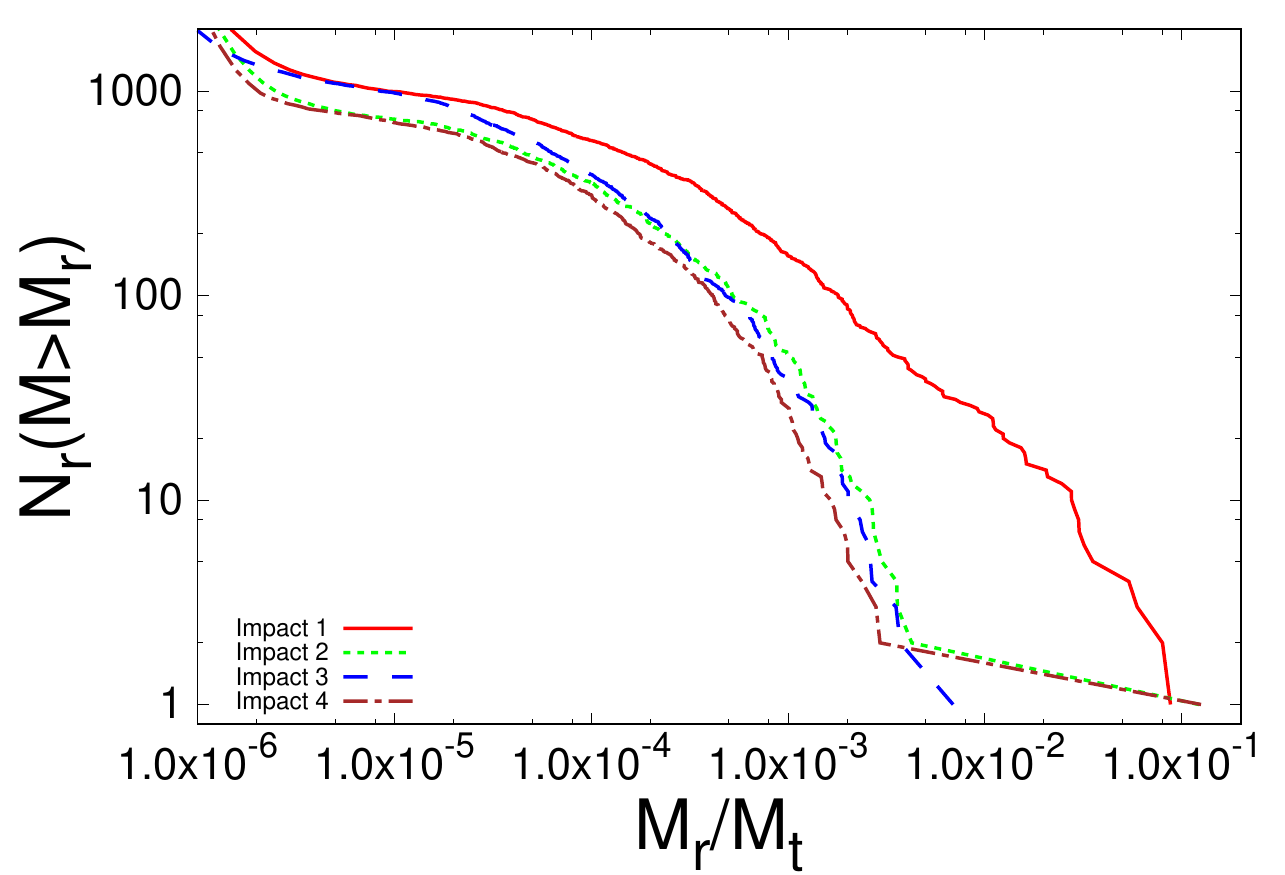}
 \caption{Cumulative mass distributions of the remnants produced through the catastrophic collisions. The horizontal axis shows the mass of the remnants $M_{{\rm r}}$ normalized by the mass of the target asteroid $M_{{\rm t}}$, the vertical axis shows the cumulative number of the remnants that have the mass larger than $M_{{\rm r}}$. The red solid curve, the green dotted curve, the blue dashed curve, and the brown chain curve show the cumulative mass distributions of the remnants produced through Impact 1, 2, 3, and 4, respectively.}
 \label{cumulative-mass-distribution-impact1-4} 
 \end{center}
\end{figure}

Figure \ref{cumulative-mass-distribution-impact1-4} shows the cumulative mass distributions of the remnants produced through the catastrophic collisions. The mass of the largest remnants are $\approx 0.1M_{{\rm t}}$ for Impact 1, 2, and 4, and $\approx 0.01M_{{\rm t}}$ for Impact 3, where $M_{{\rm t}}$ is the mass of the target asteroid. The mass distributions of the remnants show shallow slopes at remnant mass $M_{{\rm r}} \lesssim 1\times 10^{-3}M_{{\rm t}}$, indicating the lack of remnants with $M_{{\rm r}} \lesssim 1\times 10^{-4}M_{{\rm t}}$. A remnant with $1\times 10^{-4}M_{{\rm t}}$ consists of about $500$ SPH particles, with which remnants are sufficiently resolved. Therefore, the smallest size of remnants with $1\times 10^{-4}M_{{\rm t}}$ might be determined by the scale of the self gravity such as the Jeans length.

\begin{figure}[!htb]
 \begin{center}
 \includegraphics[bb=0 0 1270 680, width=1.0\linewidth]{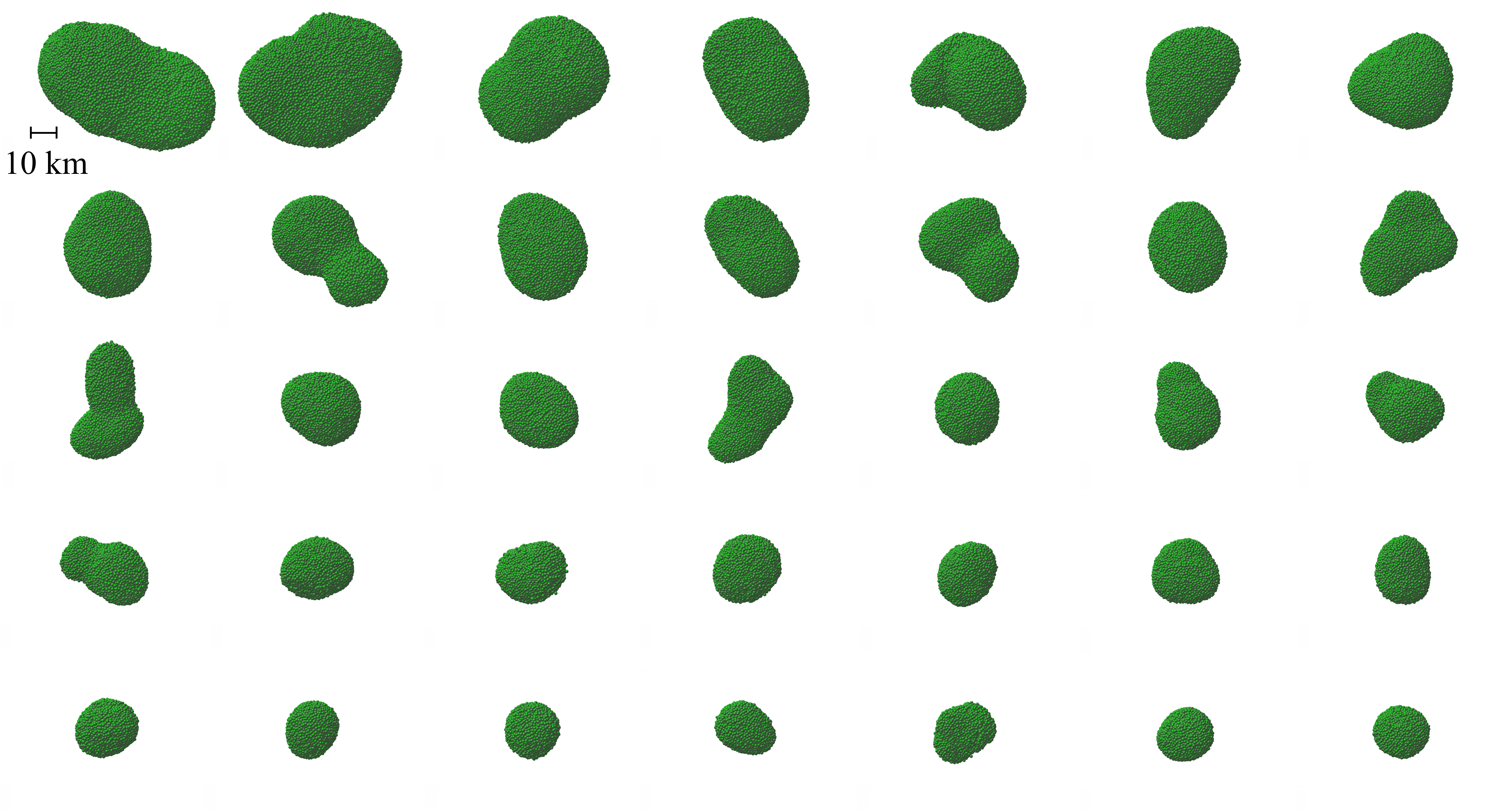}
 \caption{Remnants produced through Impact 1. This figure shows the shape of the largest remnant (the remnant at the left upper corner) to that of the 35th largest remnant (the remnant at the right bottom corner). The scale nearby the left upper remnant is valid for all the remnants in this figure.}
 \label{snapshots-of-remnants-Rt=50km-q=1.0-theta=15-v=350ms} 
 \end{center}
\end{figure}

For the analysis of remnant shapes, the well-resolved remnants that are composed of more than 5,000 SPH particles are required. We obtain 136 well-resolved remnants from Impact 1 - 4. Figure \ref{snapshots-of-remnants-Rt=50km-q=1.0-theta=15-v=350ms} shows the snapshots of major remnants produced through Impact 1. The most remnants have spherical shapes rather than elongated or flat shapes. Some remnants have characteristic shapes with two lobes. The bilobed shapes are formed through coalescence of two rounded remnants. Thus, the bilobed shapes of 67P/Churyumov-Gerasimenko visited by the spacecraft Rosetta or (486958) 2014 MU69 (nicknamed Ultima Thule) visited by the spacecraft New Horizons may be formed through catastrophic disruptions (see also \citealt{Schwartz-et-al2018}).

\begin{figure}[!htb]
 \begin{center}
 \includegraphics[width=1.0\linewidth]{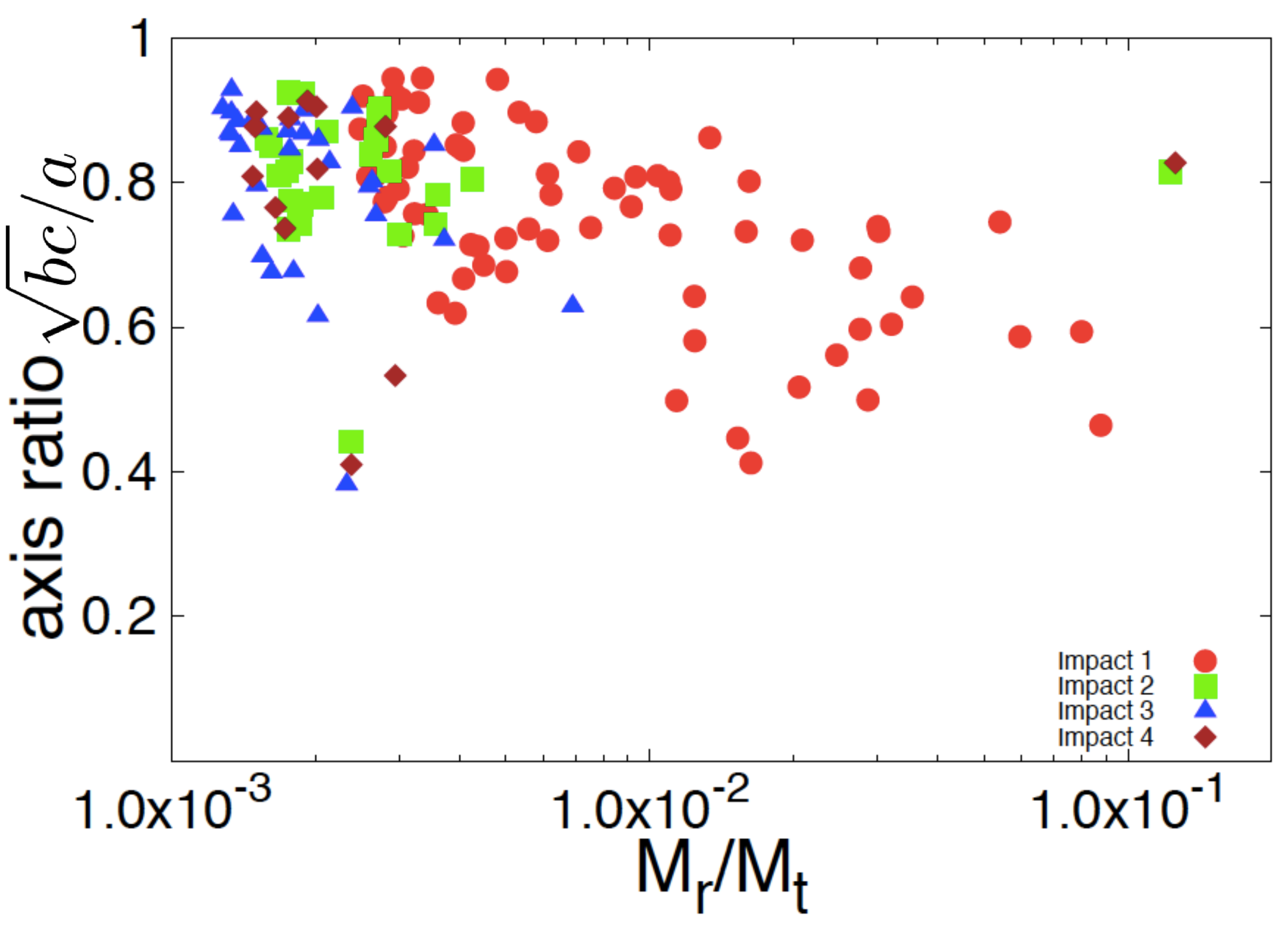}
 \caption{Axis ratio of the remnants that are composed of more than 5,000 SPH particles and are produced through the catastrophic collisions. The horizontal axis shows the mass of the remnants $M_{{\rm r}}$ normalized by the mass of the target asteroid $M_{{\rm t}}$, and the vertical axis shows $\sqrt{bc}/a$. The red circles, the green squares, the blue triangles, and the brown diamonds show the axis ratio of the remnants produced through Impact 1, 2, 3, and 4, respectively.}
 \label{summary-of-axis-ratios-of-large-clumps-impact1-4} 
 \end{center}
\end{figure}

Figure \ref{summary-of-axis-ratios-of-large-clumps-impact1-4} shows the axis ratio $\sqrt{bc}/a$ of the 136 well-resolved remnants. For the larger remnants with $M_{{\rm r}} \geq 0.01 M_{{\rm t}}$, the axis ratio $\sqrt{bc}/a$ spreads from 0.5 to 0.8. The range of the axis ratio is consistent with that of the remnants obtained from the $N$-body simulations (\citealt{Schwartz-et-al2018}). We observe the trend that the larger remnants have more irregular shapes, and this trend is also observed for the remnants with the major principal axis length $> 3\,{\rm km}$ obtained from \cite{Schwartz-et-al2018}. In contrast, the smaller remnants in our simulations mainly have $\sqrt{bc}/a \geq 0.8$, but such irregular remnants are rare in the simulations of \cite{Schwartz-et-al2018}. This difference may be caused by the difference of the numerical simulation methods or the numerical resolution.

\begin{figure}[!htb]
 \begin{center}
 \includegraphics[bb=0 0 370 262, width=1.0\linewidth]{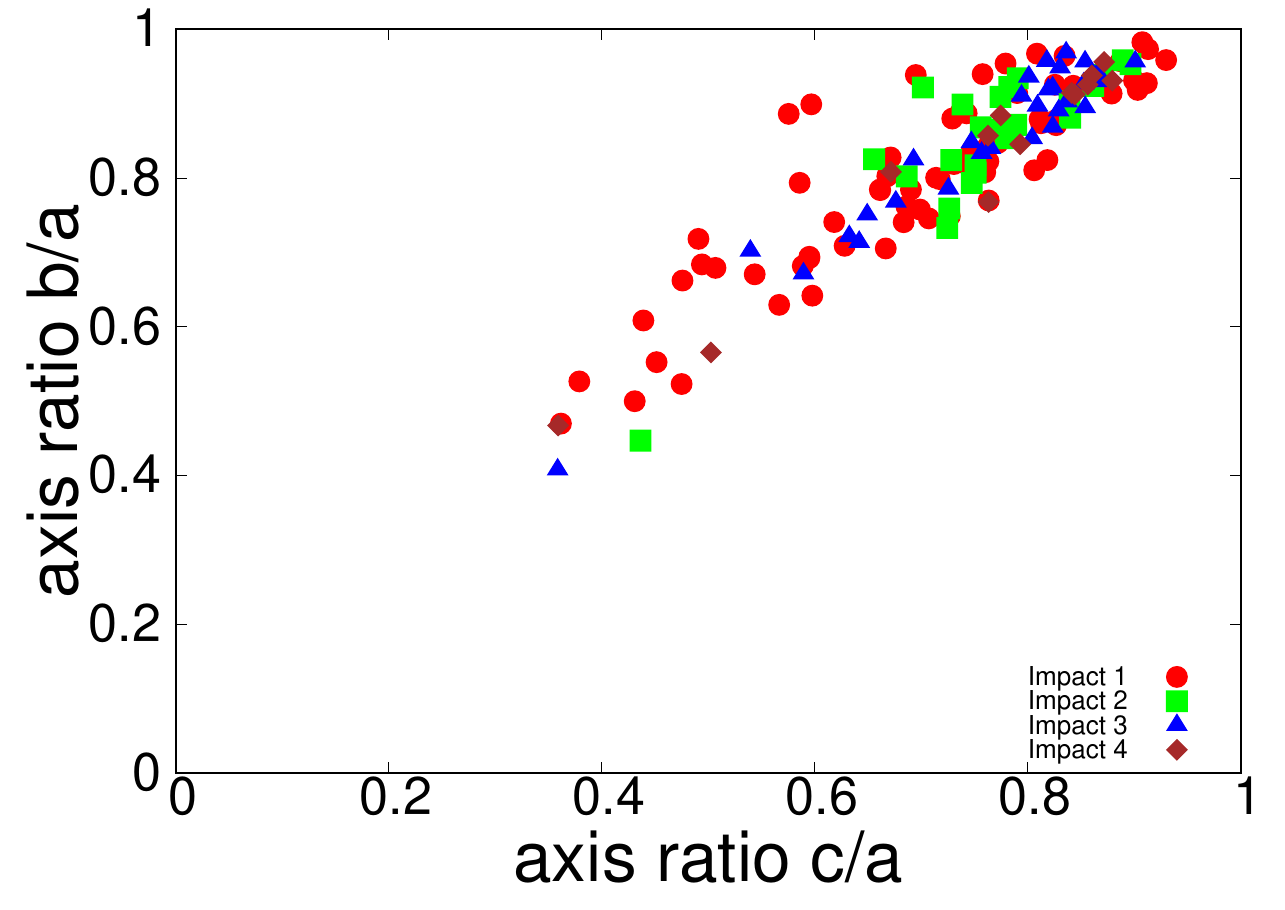}
 \caption{Shape distribution of the remnants that are composed of more than 5,000 SPH particles and are produced through the catastrophic collisions. The horizontal axis shows $c/a$ and the vertical axis shows $b/a$. Spherical shapes are plotted on the right upper side, flat shapes are plotted on the left upper side, and elongated shapes are plotted on the left bottom side. The red circles, the green squares, the blue triangles, and the brown diamonds show the axis ratios of the remnants produced through Impact 1, 2, 3, and 4, respectively.}
 \label{c_a-vs-b_a-of-large-clumps-impact1-4} 
 \end{center}
\end{figure}

Figure \ref{c_a-vs-b_a-of-large-clumps-impact1-4} shows the axis ratios of the 136 well-resolved remnants. The remnants have the axis ratios of $b/a \sim c/a$, i.e., the destructive impacts mainly produce spherical and bilobed remnants. The most elongated bilobed remnant has $b/a \approx 0.5$ and $c/a \approx 0.4$. However, there are a few flat remnants with $c/a \approx 0.6$ and there are no flat remnants with $c/a \lesssim 0.5$. Thus flat shapes are difficult to form through the catastrophic impacts.

Shapes of resultant remnants depend on the friction angle, especially for complex shapes (see \citealt{Sugiura-et-al2019}). Hence, the fraction of bilobed remnants may be affected by the choice of the friction angle. Note that the friction angle of lunar soil is $30 - 50^{\circ}$ (\citealt{Heiken-et-al1991}) and thus the friction angle of $40^{\circ}$ used in our study is the typical value.

\begin{figure}[!htb]
 \begin{center}
 \includegraphics[bb=0 0 370 272, width=1.0\linewidth]{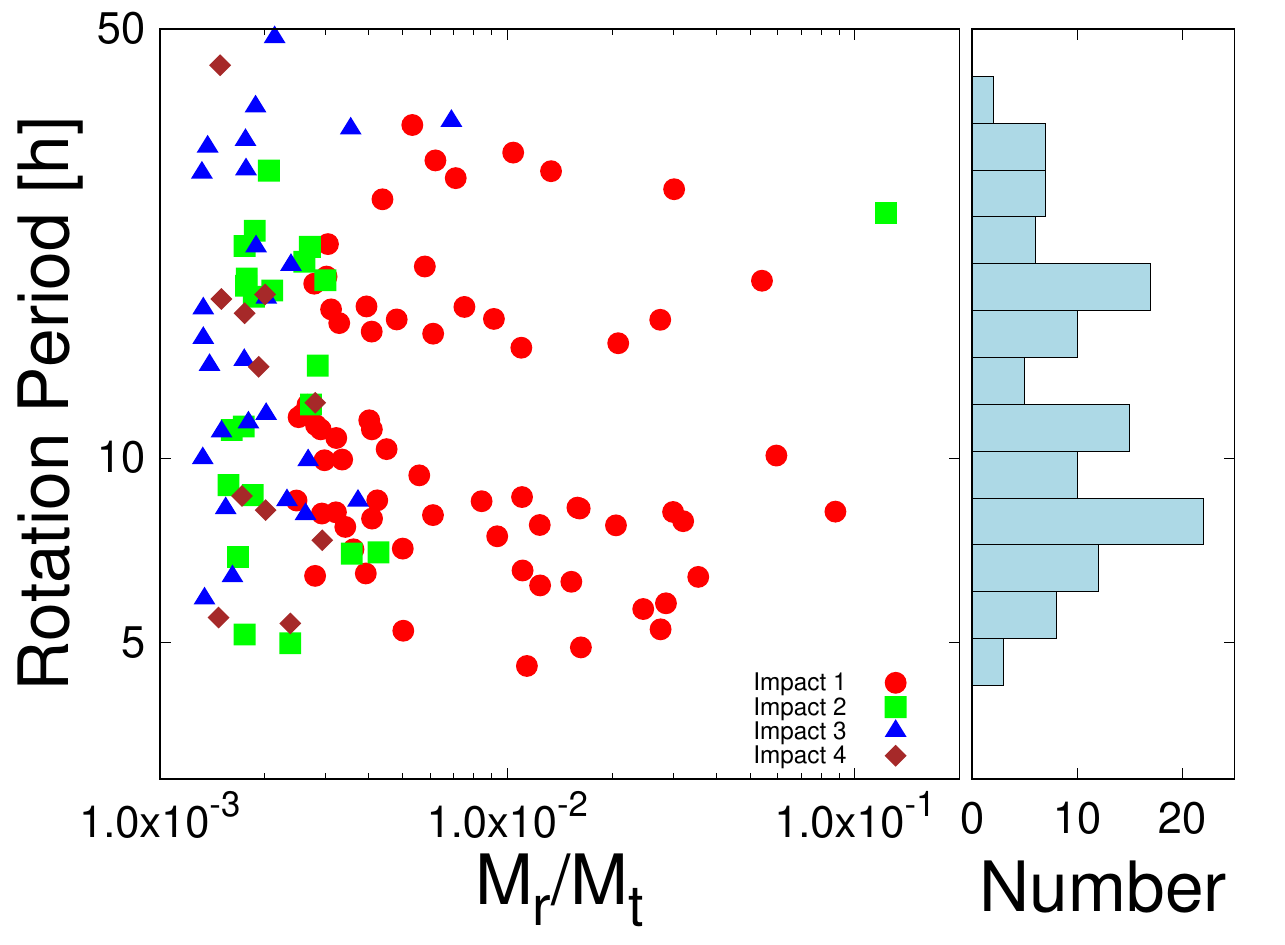}
 \caption{Rotation periods of the remnants that are composed of more than 5,000 SPH particles and are produced through the catastrophic collisions. The horizontal axis shows the mass of the remnants $M_{{\rm r}}$ normalized by the mass of the target asteroid $M_{{\rm t}}$, and the vertical axis shows the rotation periods of the remnants. The red circles, the green squares, the blue triangles, and the brown diamonds show the rotation periods of the remnants produced through Impact 1, 2, 3, and 4, respectively. Right panel shows the histogram of the rotation periods of these remnants.}
 \label{rotation-periods-of-large-clumps-impact1-4} 
 \end{center}
\end{figure}

Figure \ref{rotation-periods-of-large-clumps-impact1-4} shows the rotation periods of the 136 well-resolved remnants. The remnants have the wide range of the spin rates longer than $5\,{\rm h}$, and there is a peak at the rotation periods of about $10\,{\rm h}$. This peak rotation period is consistent with that obtained from the $N$-body simulation of the collisional destruction (\citealt{Michel-and-Richardson2013}). The remnants with the fastest rotation have the rotation periods of about $5\,{\rm h}$, which is longer than those of fast rotating asteroids such as the asteroid Bennu (period = $4.3\,{\rm h}$: \citealt{Lauretta-et-al2019}) and 1999 KW4 (period = $2.8\,{\rm h}$: \citealt{Ostro-et-al2006}). The fast rotation states may need the additional acceleration of the rotation such as the Yarkovsky-O'Keefe-Radzievskii-Paddack effect.

\section{Discussion}
The average impact velocity in the main belt is about $5\,{\rm km/s}$, so that similar-mass impacts in the present main belt mainly result in catastrophic disruptions. As we show in Section 3, it is difficult to produce flat shapes with $c/a \lesssim 0.5$ through catastrophic disruptions. Impacts with much smaller mass ratios $q$ do not result in catastrophic disruptions even in the present environment. However, such impacts probably do not produce flat shapes because deformation is considered to occur on the scale of impactors. The simulations of non-destructive impacts with $q=1/64$ and $v_{{\rm imp}}=500 - 3,000\,{\rm m/s}$ show that all the largest remnants produced through the impacts have $c/a > 0.7$ (\citealt{Sugiura2019-PhD}). Thus, we expect that flat asteroids with $c/a \lesssim 0.5$ are difficult to form in the present environment.

\cite{Sugiura-et-al2018} show that merging and nearly head-on collisions between equal-mass asteroids form flat shapes with $c/a \lesssim 0.5$. Two-body escape velocities of $100\,{\rm km}$ asteroids are about $100\,{\rm m/s}$, and impacts with impact velocities comparable to the escape velocities occur in the planet formation era (e.g.,\,\citealt{Kobayashi-et-al2016}). Thus, we suppose that flat asteroids with $c/a \lesssim 0.5$ are likely to be formed through low-velocity and non-destructive impacts in the primordial environment. The probability for nearly head-on collisions is not so high for impacts with random impact angles, and asteroids that by chance experience such nearly head-on impacts become flat shapes. This implies that asteroids that do not have flat shapes might be also primordial.

Although \cite{Bottke-et-al2005} suggest that many asteroids with the diameters $< 120\,{\rm km}$ are byproducts of disruptive collisions, \cite{Obrien-and-Greenberg2005} estimate that the timescale of collisional destruction for asteroids with the diameters $> 10\,{\rm km}$ is longer than the age of the solar system. The theoretical estimate for the fraction of primordial asteroids is uncertain. On the other hand, the classification of family asteroids using the Yarkovsky V-shapes suggests that some asteroids with the diameters $> 10\,{\rm km}$ are surviving planetesimals, i.e., they are primordial (\citealt{Delbo-et-al2019}). Thus, we suggest that flat asteroids with $c/a \lesssim 0.5$ and the diameters $> 10\,{\rm km}$ are probably primordial. In our future works, we investigate the survivability of flat asteroids in more detail because flat shapes may be easy to disrupt due to those large cross sections.

To verify our expectation, we investigate shapes of asteroids belonging to asteroid families produced through catastrophic disruptions. We consider asteroid families with $V_{{\rm lr}}/V_{{\rm tot}} < 0.5$ as families produced through catastrophic disruptions, where $V_{{\rm lr}}$ is the volume of the largest member and $V_{{\rm tot}}$ is the total volume of all the members. The volume of asteroids is calculated from those diameters. The well-resolved remnants produced through our simulation of the catastrophic disruption have sizes larger than $10\,{\rm km}$, so that we investigate asteroids with diameters larger than $10\,{\rm km}$ for the comparison. We obtain asteroid shapes from the shape model given by DAMIT database\footnote{http://astro.troja.mff.cuni.cz/projects/asteroids3D/web.php}. We also obtain information of asteroid families from AstDyS-2 database\footnote{https://newton.spacedys.com/astdys2/index.php} and diameters of asteroids from JPL small-body database\footnote{https://ssd.jpl.nasa.gov/sbdb\_query.cgi}.

\begin{figure}[!htb]
 \begin{center}
 \includegraphics[bb=0 0 370 262, width=1.0\linewidth]{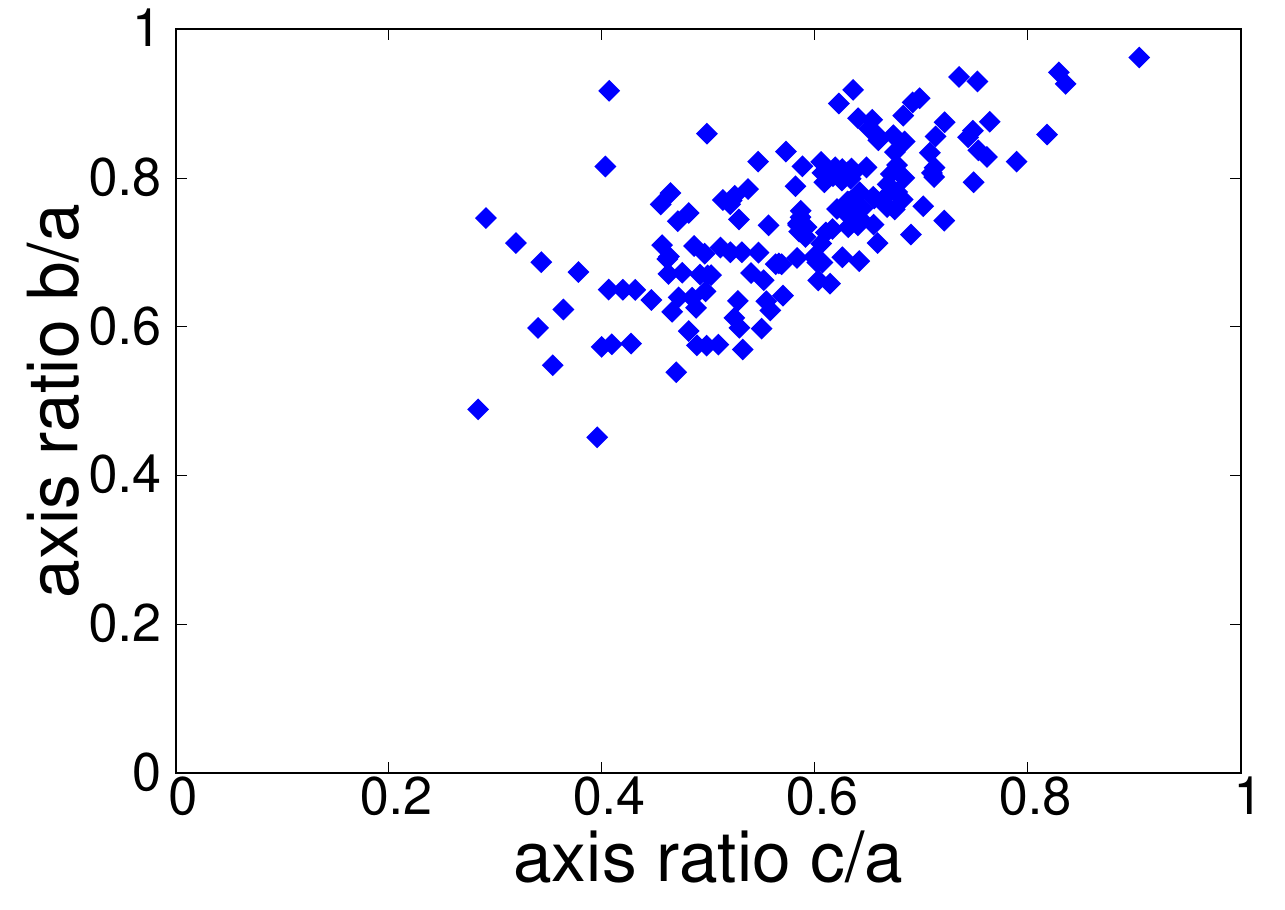}
 \caption{Shape distribution of the asteroids that belong to the asteroid families produced through catastrophic disruptions and have diameters larger than $10\,{\rm km}$. The horizontal axis shows $c/a$ and the vertical axis shows $b/a$.}
 \label{c_a-vs-b_a-of-asteroids-with-D-gt-10km-and-CD-family} 
 \end{center}
\end{figure}

Figure \ref{c_a-vs-b_a-of-asteroids-with-D-gt-10km-and-CD-family} shows the axis ratios of the asteroids belonging to the families produced through catastrophic disruptions. The axis ratios of the asteroids are mainly $b/a \sim c/a$, i.e., the asteroids mainly have spherical and elongated, or bilobed shapes. The most elongated asteroid has $b/a \approx 0.5$ and $c/a \approx 0.3$, which is consistent with the axis ratios of the most elongated remnant produced through our simulation (see Fig.\,\ref{c_a-vs-b_a-of-large-clumps-impact1-4}). However, some asteroids have flat shapes with $b/a \gtrsim 0.7$ and $c/a \lesssim 0.5$ that are not observed in our simulations.

There are some possibilities for the discrepancy regarding to flat shapes. A first possibility is observational difficulties. It is difficult to accurately determine the shapes of asteroids only by light-curve observations because there are observational errors in faint objects and uncertainties in the inversion technique. For example, the shape of (419) Aurelia was firstly estimated to be the flat shape with $b/a \approx 0.9$ and $c/a \approx 0.5$ through light-curve observations (\citealt{Hanus-et-al2016}), but later it was revised to be the spherical shape with $b/a \approx 0.9$ and $c/a \approx 0.8$ through observations with adaptive optics and occultation (\citealt{Hanus-et-al2017}).

A second possibility is interlopers. Flat asteroids belonging to families produced through catastrophic disruptions may be interlopers. Those flat shapes may be formed through low-velocity impacts in the primordial environment, and later they may be mixed up with family asteroids.

A third possibility is collisions between member asteroids. Velocity dispersion of remnants just after disruptions of parent bodies is estimated to be $\sim 100\,{\rm m/s}$ (e.g.,\,\citealt{Michel-et-al2001}), which is comparable to typical two-body escape velocities of large asteroids. Thus relative velocities between disrupted remnants are about escape velocities. They may collide with each other with low impact velocities, which may result in the formation of flat shapes if colliding two bodies have similar masses and nearly head-on collisions occur (\citealt{Sugiura-et-al2018}). A catastrophic collision ejects remnants around a plane (see Fig.\,\ref{Rt=50km-q=1.0-theta=15-v=350ms-side-view}b). Nearly head-on collisions between the remnants are reasonably probable because their orbital planes are similar (\citealt{Leleu-et-al2018}). The existence of flat asteroids belonging to asteroid families may suggest the importance of the evolution of shapes through collisions between member asteroids.

\section{Summary}
Collisional lifetimes of asteroids larger than $\sim 10\,{\rm km}$ are longer than the age of the solar system. Although the existence of asteroid families shows that some large asteroids experience catastrophic disruptions, shapes of asteroids larger than $\sim 10\,{\rm km}$ may be formed in the primordial environment and remain the same until today. Shapes of asteroids formed through catastrophic disruptions provide information on the environment of the most recent collisional events since similar-mass impacts in the present environment mainly result in catastrophic disruptions, while those in the primordial environment mainly result in non-destructive impacts.

We conduct the high-resolution simulations of the catastrophic disruptions using the SPH method for elastic dynamics with the self gravity, the fracture model, and the friction model. Our simulations show that catastrophic disruptions mainly produce spherical and bilobed shapes with $b/a \sim c/a$ (see Figs.\,\ref{snapshots-of-remnants-Rt=50km-q=1.0-theta=15-v=350ms} and \ref{c_a-vs-b_a-of-large-clumps-impact1-4}). Flat shapes with $c/a \lesssim 0.5$ are not formed through our simulations. Thus, we suggest that catastrophic disruptions are difficult to produce flat asteroids and flat asteroids are likely to be formed through low-velocity and non-destructive impacts in the primordial environment.

We investigate shapes of asteroids that have diameters larger than $10\,{\rm km}$ and belong to asteroid families produced through catastrophic disruptions. Many of them have spherical, elongated, or bilobed shapes with $b/a \sim c/a$ (see Fig.\,\ref{c_a-vs-b_a-of-asteroids-with-D-gt-10km-and-CD-family}). However, some asteroids have flat shapes with $b/a \gtrsim 0.7$ and $c/a \lesssim 0.5$, which is not consistent with the shapes of the remnants produced in our simulations of the catastrophic disruptions.

We discuss some possibilities for the discrepancy regarding to flat shapes. Flat asteroids belonging to families produced through catastrophic disruptions may be interlopers, that is, they may be formed in the primordial environment and mixed up with family asteroids. Low-velocity and similar-mass impacts between asteroids belonging to the same family may produce flat shapes, and thus the existence of flat asteroids in families may suggest the evolution of asteroidal shapes through collisions between member asteroids.

\section*{Acknowledgements}
KS is supported by JSPS KAKENHI Grant Number JP 17J01703. HK and SI are supported by Grant-in-Aid for Scientific Research (18H05436, 18H05437, 18H05438, 17K05632, 17H01105, 17H01103, 16H02160) and by JSPS Core-to-Core Program ``International Network of Planetary Science''. Numerical simulations in this work were carried out on Cray XC50 at Center for Computational Astrophysics, National Astronomical Observatory of Japan.

\section*{References}

\bibliography{mybibfile}

\end{document}